\documentclass[3p,times]{elsarticle}
\usepackage[utf8]{inputenc}
\usepackage{graphicx}
\usepackage{svg}
\usepackage{amsmath}
\usepackage{ecrc}
\usepackage[hidelinks]{hyperref}

\usepackage{subcaption}
\DeclareUnicodeCharacter{2009}{\,} 

\volume{XX}
\firstpage{1}
\journalname{}
\runauth{}
\usepackage[figuresright]{rotating}
\begin{document}
\begin{frontmatter}
\dochead{}
\title{Pulse Shape Discrimination of low-energy nuclear and electron recoils for improved particle identification in NaI:Tl}
\author[a,b]{N.~J.~Spinks\corref{cor1}}
\cortext[cor1]{Corresponding author}
\ead{nathan.spinks@anu.edu.au}
\author[a,b]{L.~J.~Bignell\corref{cor1}}
\ead{lindsey.bignell@anu.edu.au}
\author[a,b]{G.~J.~Lane}
\author[b]{A.~Akber}
\author[a,c]{E.~Barberio}
\author[a,c]{T.~Baroncelli}
\author[b]{B.~J.~Coombes}
\author[b]{J.~T.~H.~Dowie}
\author[b]{T.~K.~Eriksen}
\author[a,b]{M.~S.~M.~Gerathy}
\author[b]{T.~J.~Gray}
\author[a,c]{I.~Mahmood}
\author[b]{B.~P.~McCormick}
\author[a,c]{W.~J.~D.~Melbourne}
\author[b]{A.~J.~Mitchell}
\author[a,c]{F.~Nuti}
\author[b,d]{M.~S.~Rahman}
\author[a,c]{F.~Scutti}
\author[a,b]{A.~E.~Stuchbery}
\author[d]{H.~Timmers}
\author[a,c]{P.~Urquijo}
\author[a,b]{Y.~Y.~Zhong}
\author[a,c]{M.~J.~Zurowski}

\address[a]{ARC Centre of Excellence for Dark Matter Particle Physics}
\address[b]{Department of Nuclear Physics and Accelerator Applications, The Australian National University, Canberra, ACT 2601, Australia}
\address[c]{School of Physics, University of Melbourne, Melbourne, Victoria 3010, Australia}
\address[d]{School of Science, University of New South Wales, Canberra, ACT 2610, Australia}
\begin{abstract}
 The scintillation mechanism in NaI:Tl crystals produces different pulse shapes that are dependent on the incoming particle type. The time distribution of scintillation light from nuclear recoil events decays faster than for electron recoil events and this difference can be categorised using various Pulse Shape Discrimination (PSD) techniques. In this study, we measured nuclear and electron recoils in a NaI:Tl crystal, with electron equivalent energies between $2$ and $40$~keV. We report on a new PSD approach, based on an event-type likelihood; this outperforms the charge-weighted mean-time, which is the conventional metric for PSD in NaI:Tl. Furthermore, we show that a linear combination of the two methods improves the discrimination power at these energies.

\end{abstract}
\end{frontmatter}

\section{Introduction}
One of the greatest unsolved mysteries of modern physics is the nature of dark matter \cite{ade2016planck}, with Weakly Interacting Massive Particles (WIMPs) being a strongly motivated dark matter candidate. WIMPs are expected to interact with nuclei, resulting in nuclear recoil events. Direct-detection experiments aim to search for WIMP signals by measuring the energy deposition of nuclear recoils in a detector, in excess of known backgrounds such as electron recoils. To date, standard WIMP assumptions have been ruled out by most direct-detection experiments~\cite{schumann2019direct}, with the exception of DAMA/LIBRA, which has measured an annual modulation signal with a high statistical significance of 12.9$\sigma$ that is compatible with dark matter~\cite{bernabei2018first}. A key difference between DAMA/LIBRA and these other experiments is the use of NaI:Tl crystals as a detection medium. This has motivated other NaI:Tl-based direct-detection experiments~\cite{antonello2019sabre, Amar__2021, adhikari2018initial}, and highlights the necessity of a model-independent test of DAMA/LIBRA. 

Pulse Shape Discrimination (PSD) is a common approach used to identify the type of particle interacting in a detector, and it generally relies on the fact that the shape of a recorded signal changes for different particle types. PSD can be utilised in NaI:Tl to distinguish between nuclear and electron recoils, based on the scintillation response, which depends on the population and de-population of the excited states of the crystal lattice. Absorption of energy can promote an electron to the conduction band to produce an exciton which may then decay to the ground state and generate an optical photon. The time distribution of the emitted photons exhibits components that include a rise time and a `fast' decay time of 250~ns~\cite{birks1954scintillation}. Additionally, a `slow' component of scintillation light is observed in NaI:Tl due to excited states that are forbidden from decaying to the ground state. The proportion of scintillation light that goes to this slow component ($\tau$=1.5 $\mu$s in NaI:Tl~\cite{jbbirks}) is dependent on the stopping power for the incident radiation. Since different types of particles have characteristic stopping powers, it is possible to differentiate between nuclear recoil and electron recoil events in NaI:Tl from the differences in their pulse shapes~\cite{knollrad}.

As backgrounds in NaI:Tl are predominantly electron recoils, PSD is a powerful tool for background reduction in these detectors, with analyses primarily limited to a charge-weighted mean-decay-time algorithm~\cite{lee2015pulse}. COSINE and ANAIS, two NaI:Tl-based dark matter direct-detection experiments, have considered analysis of pulse shapes, albeit with a focus on selection of PMT noise events~\cite{Adhikari_2021, antonello2021characterization} or identification of muon interactions~\cite{Amar__2021}, rather than for the classification of different recoil types. PSD has also been applied to other materials such as liquid xenon for $\gamma$-ray background suppression~\cite{akimov2002measurements}, leading to the development of methods other than the mean-decay-time, for example a maximum likelihood method for liquid argon~\cite{lippincott2008scintillation}. 

In this paper we present measurements of scintillation events in a NaI:Tl crystal, including methods to classify events as being due to either nuclear or electron recoils in an energy range relevant for WIMP dark matter searches. We present a comparison of the particle discrimination power between a conventional charge-weighted mean-decay-time algorithm and a log-likelihood ratio approach, and also determine a linear combination of these two techniques that improves the discrimination power beyond what can be achieved with either on its own.

\section{Experiment Design}
\begin{figure}[ht]
    \centering
    \includegraphics[scale=0.4]{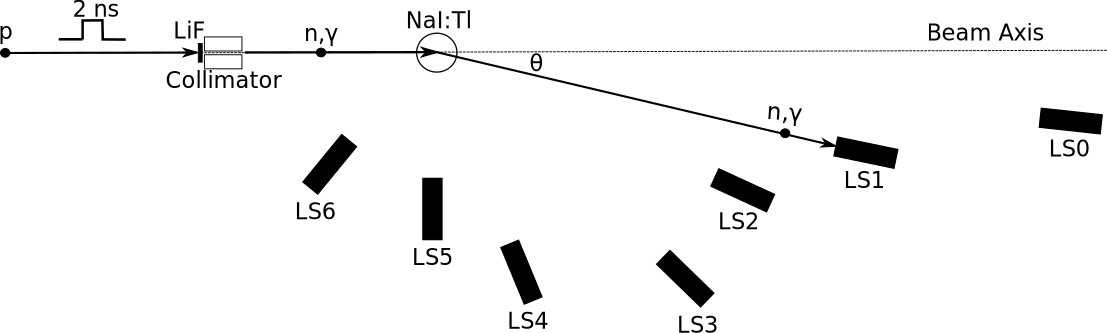}
    \caption{Schematic of the experimental setup. A 3-MeV beam of bunched, pulsed protons hits a tantalum-backed LiF target. Neutrons resulting from $^7$Li(p,n)$^7$Be reactions are collimated to interact with the NaI:Tl crystal, scattering into one of seven EJ-309 liquid scintillator detectors, labelled LS0-LS6.}
    \label{fig:Exp_geometry}
\end{figure}

A set of characteristic signals in NaI:Tl corresponding to nuclear and electron recoil events was obtained as part of experiments aimed at measuring the quenching factor in NaI:Tl. Complete experimental details can be found in Ref.~\cite{Bignell_2021}; this section provides an overview.  

A schematic diagram of the experimental setup is shown in Figure~\ref{fig:Exp_geometry}. A proton beam with an energy of 3 MeV was delivered by the 14UD electrostatic tandem accelerator at the Australian National University's Heavy Ion Accelerator Facility. The protons were incident on a 520-$\mu$g/cm$^2$ thick LiF target, producing a beam of quasi-monoenergetic neutrons through the $^7$Li(p,n)$^7$Be reaction. A tantalum backing of thickness 200~mg/cm$^2$ was used to stop the proton beam. The beam was pulsed with bursts of FWHM $<$2~ns arriving every 747~ns, and the resulting neutrons were collimated onto the NaI:Tl crystal. Scintillator detectors were used to detect neutrons scattered from the crystal. 

Seven EJ-309 liquid scintillator detectors were placed at angles between 12.5$^{\circ}$ and 135$^{\circ}$, relative to the NaI:Tl crystal and with respect to the beam axis, and placed at distances of 200 mm to 609 mm. These detectors are of length 15 cm and diameter 38 mm and were designed to achieve high scattering cross-sections while maintaining acceptably low neutron transit times of $<$15~ns across the detector. EJ-309 was selected as the scintillator due to its excellent particle identification capablility and high light yield. The scintillation light was registered via 1.5-inch diameter Hamamatsu H10828 ultra-bialkali photomultipliers. 

The cylindrical NaI:Tl crystal of height 40 mm and diameter 40 mm was packaged inside an aluminium enclosure with a borosilicate glass window; two Hamamatsu H11934-200 1x1 inch ultra-bialkali photomultipliers were coupled to the window of the crystal enclosure with optical grease and wrapped with PTFE and electrical tape. This crystal, manufactured in 1988 with an unknown growth method and thallium concentration, had a measured light yield of $1.8$ photoelectrons per keV (PE/keV), determined by fitting a Gaussian to the single-photoelectron peak of a $^{241}$Am calibration source. This light yield is considered to be quite low, with the light yield from high quality NaI:Tl crystals expected to be on the order of 14.0 to 15.5 PE/keV, as seen in the initial performance of the COSINE-100 experiment~\cite{adhikari2018initial}. The cause of this low light yield can be attributed to both poor optical coupling, with square PMTs fit on a smaller, round crystal, and also a likely low intrinsic light yield from the NaI:Tl crystal.

\section{Data Processing and Event Selection}
Waveforms from all nine detector channels --- one PMT for each of the seven liquid scintillator detectors, along with the two PMTs on the NaI:Tl detector --- were acquired using an XIA Pixie-16 digitiser with a 500-MSPS sampling rate and 12-bit resolution. A beam-sync signal was produced in phase with the RF signal that drives the accelerator beam pulsing system and was recorded as a time relative to the digitiser clock. Additionally, a BaF$_2$ detector was placed at a backwards angle to record a timing signal, corresponding to prompt $\gamma$ rays emitted when the pulsed beam hit the target. This timing signal was used to correct for the phase drift between the timing signal of the beam and the times of recorded events at the target. Each channel was digitised when the signal exceeded a threshold that was set above the noise pedestal, but below a single photoelectron level, with the data acquisition requiring coincident signals from at least one of the liquid scintillator detectors and a NaI:Tl channel to trigger.

Waveforms were processed to determine key characteristics such as the pulse arrival time and the total charge. Understanding these two signal properties allows for nuclear and electron recoil events to be identified and selected from the overall dataset. The pulse arrival time was evaluated by setting a threshold at 30$\%$ of the maximum waveform amplitude and determining the time at which the waveform first crossed this threshold.

The charge of the waveforms recorded from the PMTs of the NaI:Tl detectors was determined by an integration algorithm. To avoid integrating noise between separated single-photoelectron events, regions where the waveform charge was in excess of a threshold of half the average single-photoelectron pulse height were selected. These regions were extended by two samples (4~ns) either side of where the threshold crossing occurred, to account for the rise and fall time of the signal. For events in the liquid scintillator detectors, charge was integrated for two distinct time windows: fast charge, $Q_{fast}$, corresponded to the first 10~ns after the pulse arrival time, and tail charge, $Q_{tail}$, the total charge after the first 10~ns. The total charge of the waveform is taken as the sum of these two components. For further details, including a schematic representation of this approach, see Ref~\cite{Bignell_2021}.

\begin{figure}[ht]
    \centering
    \includegraphics[width=\textwidth]{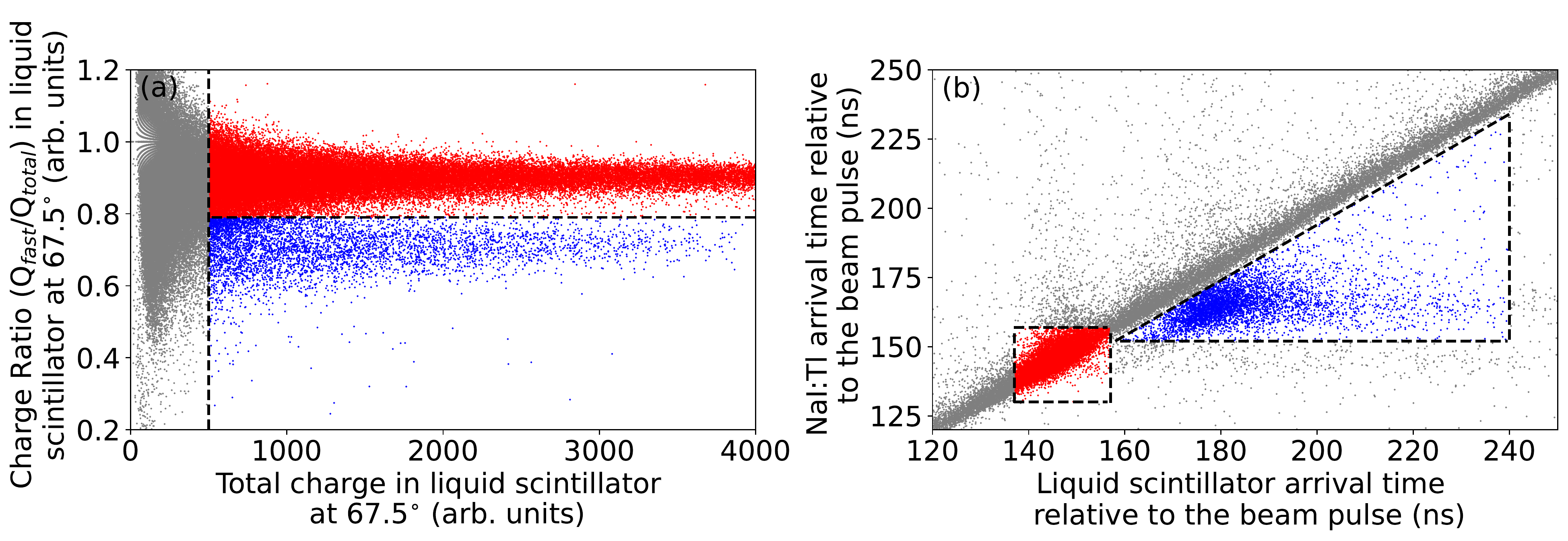}
    \caption{The data in both panels corresponds to the liquid scintillator at 67.5$^{\circ}$ to the beam axis. (a) Total pulse charge versus the ratio of the charge in the fast component of the pulse to the total pulse charge. Two distinct bands are produced, with the band at higher charge ratios due to electron recoils, while the lower band is the result of nuclear recoils. (b) The time difference between the signal arrival in the NaI:Tl detector and the beam pulse versus the time difference between the signal arrival in the liquid scintillator detector and the beam pulse. Nuclear recoils appear $\approx$20~ns later in the NaI:Tl crystal, compared to electron recoils, and $\approx$40~ns later in the liquid scintillator; this is consistent with kinematics. In both cases, cuts are selected for nuclear recoil events (blue) and electron recoil events (red). Events which cannot be clearly selected as either event type are in gray.}
    \label{fig:TOF_Q_cuts}
\end{figure}

These two waveform properties, $Q_{fast}$ and $Q_{tail}$, are used for particle identification in the liquid scintillator detectors. Events are selected as either neutrons or gammas by considering the calculated charge ratio values $Q_{fast}$/$(Q_{fast}+Q_{tail})$. These values are shown in Figure \ref{fig:TOF_Q_cuts}(a); there are two distinct bands that correspond to neutron and $\gamma$-ray events in the liquid scintillator. Neutrons (blue) were identified as events that had a lower charge ratio than gamma events (red). Events with low total charge (gray) are excluded since there is no clear separation between the neutron and gamma events. Additionally, the rejection of these low-charge events minimises the probability that events selected using this charge ratio cut are misclassified. Furthermore, we also remove low-energy noise-like events in the NaI:Tl detector by retaining only events that pass a charge asymmetry criterion of:
\begin{equation}
    \frac{|Q_1 - Q_2|}{Q_1 + Q_2} < 0.5,
\end{equation} 
where $Q_1$ and $Q_2$ are the measured charge from each of the NaI:Tl PMTs. 

From the arrival time of the liquid scintillator detector signal relative to the arrival time of the NaI:Tl signal, events can be selected based on kinematics, recognising that neutrons have a longer time-of-flight compared to $\gamma$ rays. An example timing-timing coincidence plot is shown in Figure \ref{fig:TOF_Q_cuts}(b). The red events are prompt $\gamma-\gamma$ coincidence events, since they are earliest in time. These events correspond to electron recoil events in the NaI:Tl scintillator detector since the timing implies they are either photons scattered from the NaI:Tl crystal into the liquid scintillator detectors, or two $\gamma$ rays created by the proton beam hitting the target that then hit the two detectors. 

The diagonal feature in Figure \ref{fig:TOF_Q_cuts}(b) is comprised of fast coincident events hitting both the NaI:Tl detector and a liquid scintillator detector. These are events such as Compton scattering between two detectors or cosmic showers triggering both detectors and are rejected in the event-selection process. The blue events correspond to neutron induced nuclear recoil events, which are delayed in both detectors relative to the $\gamma-\gamma$ events.

\begin{figure}[ht]
    \centering
    \includegraphics[scale=0.4]{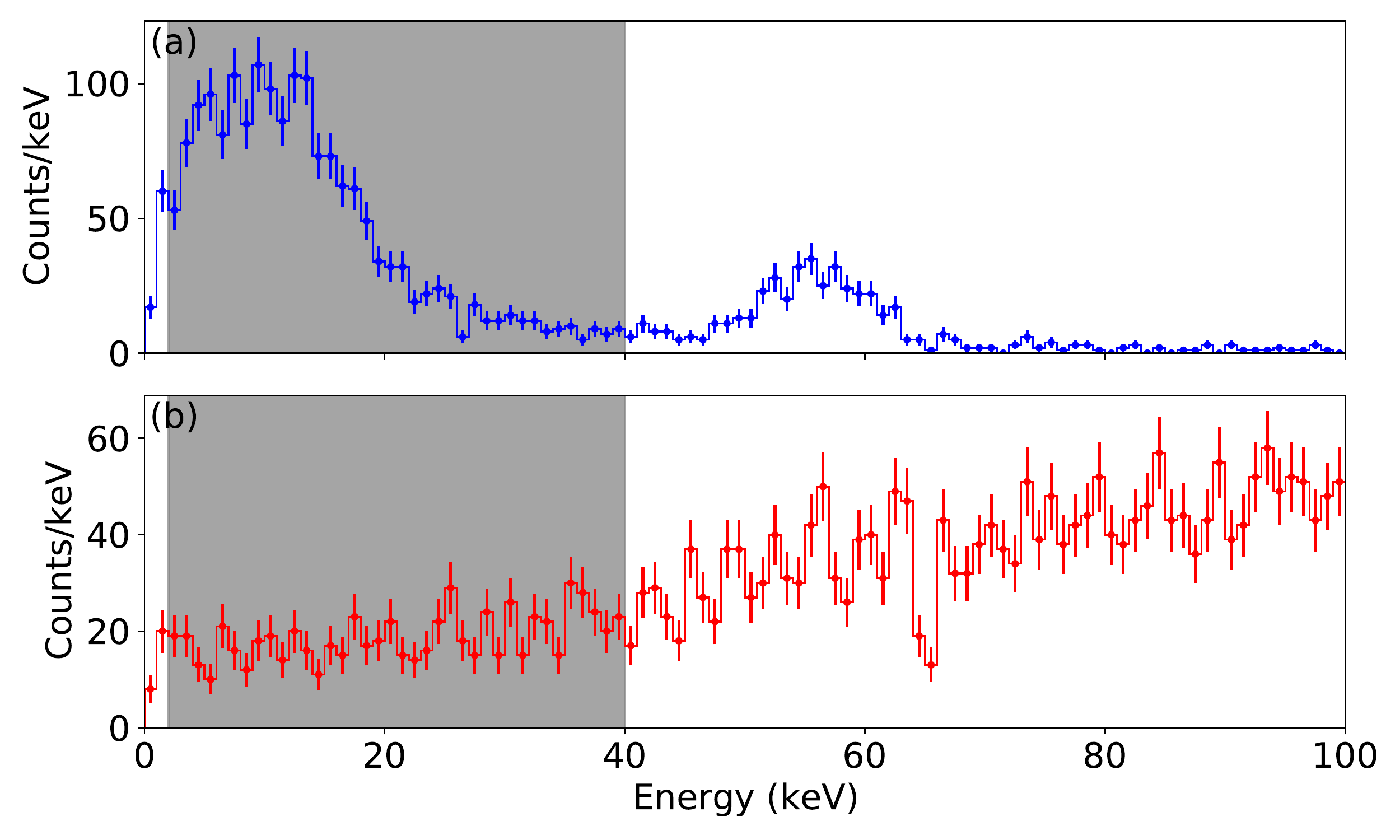}
    \caption{(a) The energy spectrum of neutron events in NaI:Tl, in coincidence with the 67.5$^{\circ}$ liquid scintillator detector, after timing and charge ratio cuts are applied. The distinct peak at 57.6~keV is due to the de-excitation of the first excited state in $^{127}$I. A gray band is used to highlight the $2-40$~keV cut placed on the spectrum to produce a nuclear recoil tagged dataset. (b) The energy spectrum of $\gamma-\gamma$ events in NaI:Tl, in coincidence with the 67.5$^{\circ}$ liquid scintillator detector, after timing and charge ratio cuts are applied. The same energy cut as was used for a) is applied to this spectrum to produce an electron recoil tagged dataset.}
    \label{fig:Energy_Spectra}
\end{figure}

After cuts are applied, two subsets of data are selected, based on particle type. It should be noted that the integrated charge of these events, as shown in Figure \ref{fig:TOF_Q_cuts}, is in arbitrary units that must be calibrated to keV. The energy scale for the charge was calibrated with $^{137}$Cs, $^{241}$Am and $^{133}$Ba $\gamma$-ray sources. The peaks were fit using a Gaussian-on-linear-background model. The spectra from these calibration sources were corrected by the known non-linear electron recoil response of NaI:Tl ~\cite{khodyuk2010nonproportional, valentine1994design} after fitting. Each peak corresponds to a point on an energy-charge plot, allowing for a linear fit to be applied and produce a calibration factor from the fit gradient. As this charge-to-energy unit conversion is determined through the measurement of gamma-sources, the energy scale provided in keV is an electron-equivalent energy in all cases. 

The energy spectrum of events associated with neutrons measured in the NaI:Tl detector, in coincidence with the 67.5$^{\circ}$ liquid scintillator detector, consists of two prominent peaks (see Figure \ref{fig:Energy_Spectra}(a)). The lower-energy peak is attributed to recoiling Na nuclei, determined from the known system kinematics (detector angle and distance, beam energy, target thickness, etc). The higher-energy peak corresponds to the de-excitation of a 57.6-keV excited state of $^{127}$I. This de-excitation peak is rejected by selecting only events that have energies from $2-40$~keV. The energy spectrum of events associated with $\gamma$-rays in the NaI:Tl detector does not have distinct peaks (see Figure \ref{fig:Energy_Spectra}(b)), however the same $2-40$~keV cut is placed on this spectrum to retain only electron recoil events and to enable an energy-dependent study of the two recoil types. This process was repeated for each NaI:Tl and EJ-309 detector combination, before all the data subsets were combined into one nuclear recoil dataset and one electron recoil dataset, containing $6838$ and $5843$ events, respectively (see Table \ref{table:NR_ER_stats}). 

\begin{table}
\centering
\begin{tabular}{|c|c|c|}
  \hline
  Energy (keV)&
  $\#$ Nuclear recoils &
  $\#$ Electron recoils \\
  \hline
  $2-5$   & $1255$ & $401$ \\
  $5-10$  & $1431$ & $604$ \\
  $10-15$ & $1225$ & $633$ \\
  $15-20$ & $945$ & $731$ \\
  $20-25$ & $661$ & $801$ \\
  $25-30$ & $540$ & $815$ \\
  $30-35$ & $425$ & $886$ \\
  $35-40$ & $356$ & $972$ \\
  \hline
  $2-40$ & $6838$ & $5843$ \\
  \hline
\end{tabular}
  \caption{Number of nuclear and electron recoil events in each studied energy bin after the event selection process, with the total statistics over the $2-40$~keV energy range also provided.}
  \label{table:NR_ER_stats}
\end{table}

\section{Results and Discussion}
After event selection, the noise in each waveform was suppressed. Signal amplitudes in a region before the pulse arrival time are used to produce a noise histogram. We deduced a 3$\sigma$ noise rejection threshold based on this distribution, and waveform amplitudes below this threshold were treated as noise and set to zero. Each noise-suppressed waveform was then normalised to an area of one. Additionally, each waveform was shifted to align the pulse arrival times to a common value.  

\begin{figure}[ht]
    \centering
    \includegraphics[width=\textwidth]{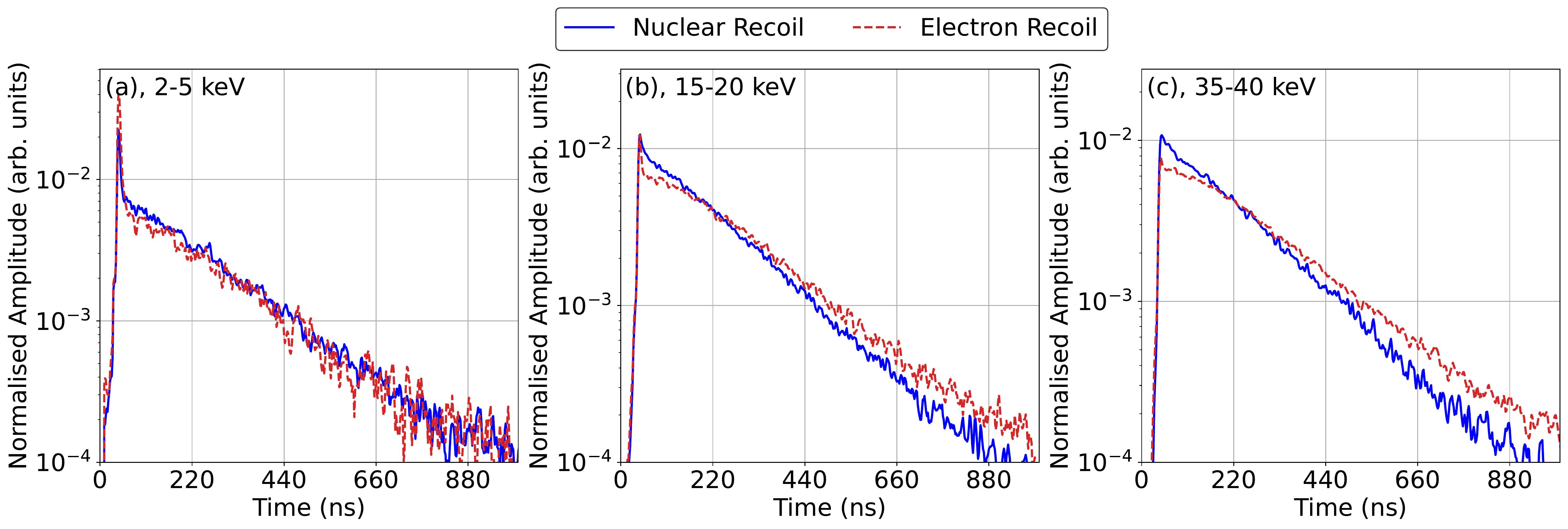}
    \caption{Normalised, averaged pulse shapes for tagged nuclear recoil (blue) and electron recoil events (red) extracted from the data, as described in the text. Waveforms are shown for the energy bins (a) $2-5$~keV (b) $15-20$~keV and (c) $35-40$~keV. In each case, the nuclear recoil events can be seen to decay faster than electron recoils, which is expected from the scintillation mechanism in NaI:Tl~\cite{knollrad}.}
    \label{fig:Tagged_Average}
\end{figure}

As a preliminary check of our selection process, averaged waveforms were produced to assess the consistency of the results. The average waveforms in the $2-5$~keV, $15-20$~keV and $35-40$~keV energy bins are shown in Figure \ref{fig:Tagged_Average}. Waveforms tagged as nuclear recoils (NR) are seen to decay faster than the electron recoils (ER), as expected from the scintillation mechanism in NaI:Tl~\cite{knollrad}. The averaged pulse has a distinct spike at $\approx$40~ns, a result of requiring an initial photoelectron peak to trigger event collection. This spike is higher at low energies because the initial spike always occurs in the same time position, and is a larger fraction of the total signal. 

Signals produced by nuclear and electron recoils in the crystal have different time distributions. These distributions can be characterised by applying suitable transformations to the waveform data that produce a metric capable of discrimination between the different particle types. For the three techniques outlined in this section, waveform data are analysed in $5$-keV energy bins for both nuclear and electron recoil events, shown in blue and red respectively, with the exception of the $2-5$-keV bin. This analysis is performed on data for all of the eight energy bins listed in Table \ref{table:NR_ER_stats}, however detailed figures are only provided here (for each PSD approach) for the same three energy bins shown in Figure \ref{fig:Tagged_Average}, highlighting the energy dependence of the separation between recoil types. Results for all bins spanning the $2-40$-keV energy region are summarised at the end of this section. The primary reason for focusing on analysis of the $2-40$-keV region is due to the low light yield of 1.8 PE/keV. NaI:Tl crystals in current dark matter direct-detection experiments generate 5 to 10 times more photoelectrons per keV of energy~\cite{adhikari2018initial}. Waveforms up to 20~keV in the present data have a similar number of photoelectrons as events in the $2-6$-keV DAMA energy region for these other experiments, allowing us to predict the discrimination power for a higher-quality crystal will be in the WIMP region of interest.

\subsection{Charge-Weighted Mean-Time (lnMT)}
\begin{figure}[ht]
    \centering
    \includegraphics[width=\textwidth]{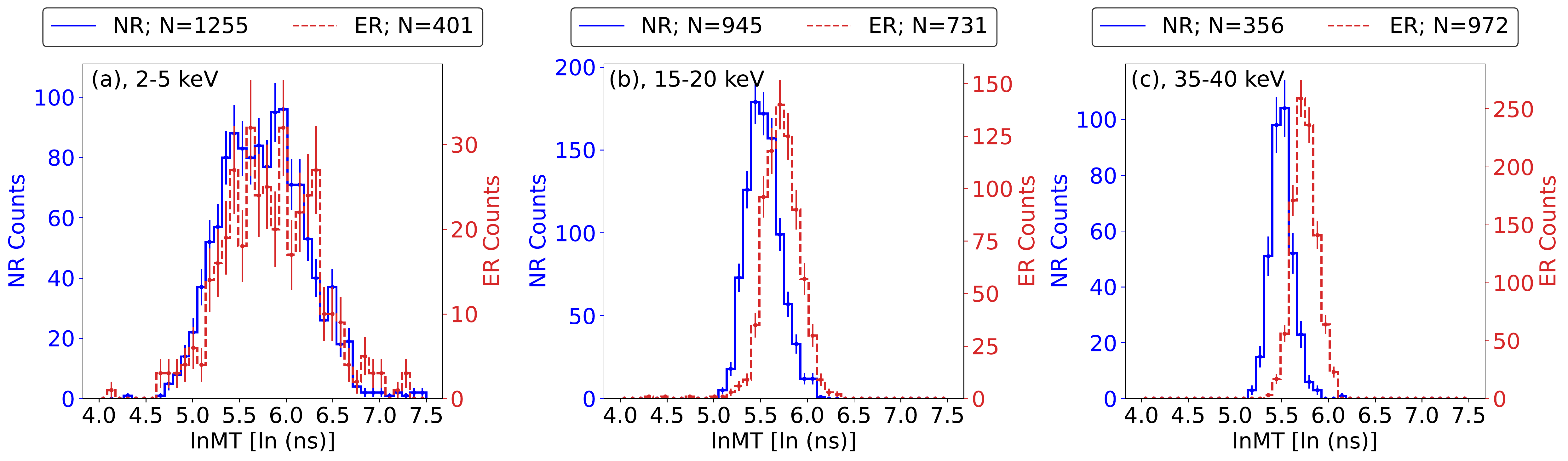}
    \caption{lnMT distributions for the tagged nuclear recoil (blue) and electron recoil (red) data sets, binned in energy ranges of (a) $2-5$~keV, (b) $15-20$~keV, (c) $35-40$~keV. On average, tagged electron recoils have a longer mean decay time than nuclear recoil events.}
    \label{fig:lnMT_dists}
\end{figure}

We reproduce the approach taken in previous work~\cite{lee2015pulse} by computing the natural log of the charge-weighted mean-time for each waveform, defined by:
\begin{equation}
\ln{\left(\text{MT}\right)} = \ln{\left(\frac{\sum_{i=1}^k Q_i t_i}{\sum_{i=1}^k Q_i}\right)},
\end{equation} where $Q_i$ is the charge and $t_i$ is the time of the $i$th point on a waveform, comprised of $k$ points. Figure \ref{fig:lnMT_dists} shows binned data when the charge-weighted mean-time (lnMT) method is applied to the tagged time-distribution data, with uncertainties and number of events provided for each histogram. The distributions are quite distinct for all energies greater than $5$~keV, with larger separation observed at higher energies.

\begin{figure}[ht]
    \centering
    \includegraphics[scale=0.33]{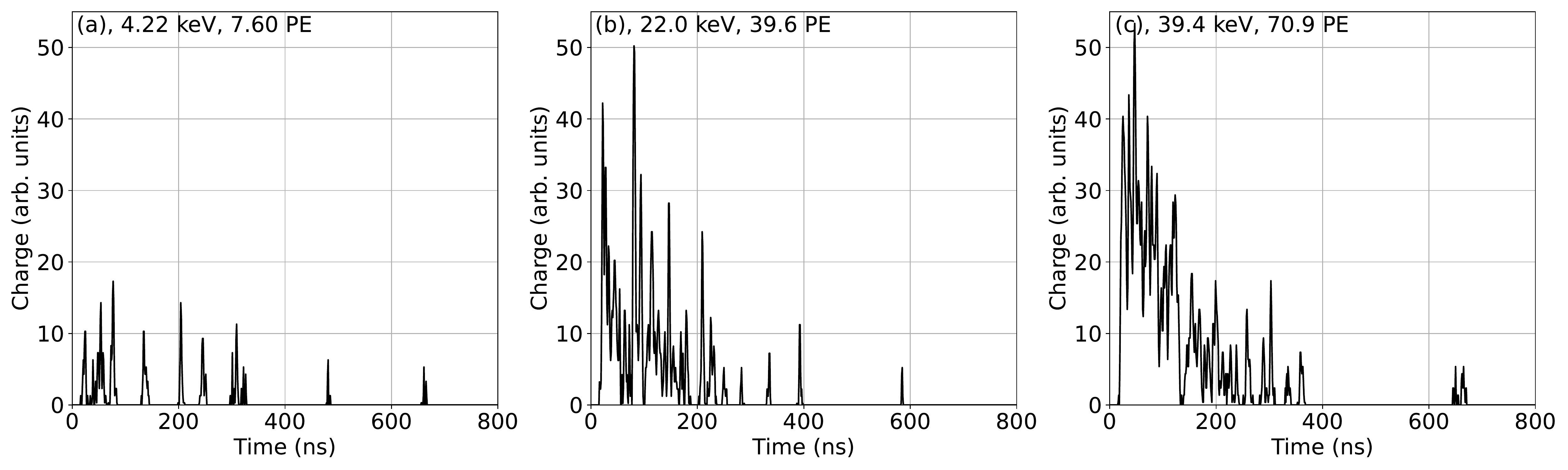}
    \caption{Digitised, noise-suppressed, signals from a PMT connected to a NaI:Tl crystal, with a low light-yield of 1.8 PE/keV. Higher-energy events consist of more photoelectrons, producing a signal with a more well-defined pulse shape. In each example provided, the waveforms are noise-suppressed and aligned in time. (a) A 4.22-keV waveform consisting of sparse peaks due to the low light-yield. (b) A 22.0-keV waveform. With an anticipated 5-10 times greater light-yield, this would be representative of pulses expected in the region of interest to a dark matter experiment, in terms of the numbers of photoelectrons. (c) A 39.4-keV waveform, where more than 70 photoelectrons are detected and the pulse shape is better defined.}
    \label{fig:light_yield}
\end{figure}

Despite a significant reduction in separation at low energies, it retains some classification power. The discrimination power at low energies is considered to be poor, and would be expected to improve for a crystal with a light yield than the present 1.8 PE/keV. Example waveforms of energies 4.22~keV, 22.0~keV and 39.4~keV are provided in Figure \ref{fig:light_yield}, and highlight the effect a higher number of photoelectrons has on the overall pulse shape. The low-energy waveforms are comprised of few photoelectrons, and the mean-decay-time is susceptible to large statistical variance. 

\subsection{Log-Likelihood Ratio}
\begin{figure}[ht]
    \centering
    \includegraphics[scale=0.44]{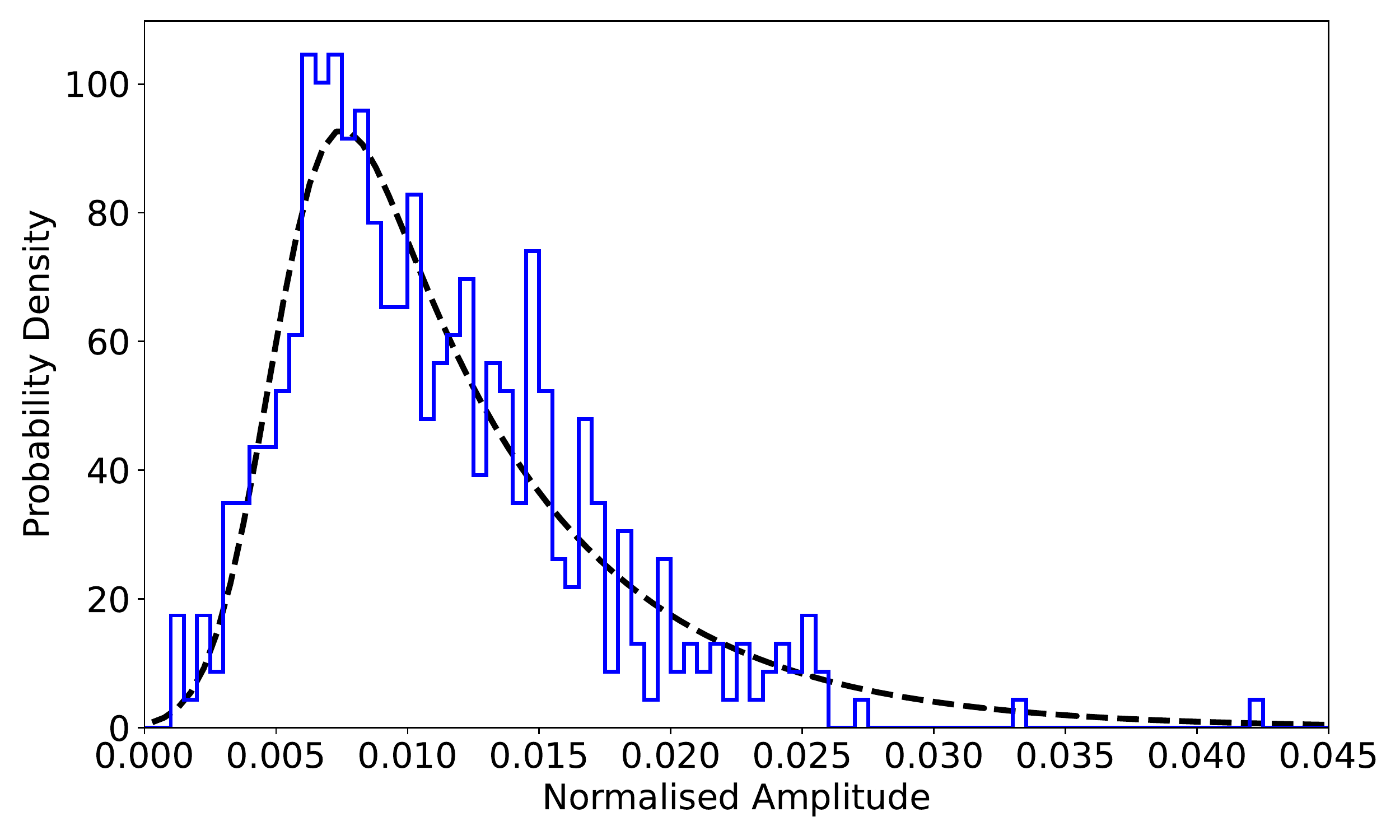} 
    \caption{The probability density function of waveform amplitudes 10~ns after pulse triggering for all nuclear recoil events in the $15-20$-keV bin. The data are fit with an exponentially modified Gaussian and the area under this fit is normalised to one. Similar PDFs are created for every time bin in the waveform.}
    \label{fig:PDF_example}
\end{figure}

The log-likelihood ratio (LLR) function is a probabilistic approach for quantifying how likely it is that a waveform is due to a nuclear recoil. By using this approach, we use information provided by all data in our tagged datasets every time a LLR is calculated, rather than the event-by-event calculation from the lnMT metric. In doing so, we aim to provide a higher-quality discrimination approach. The LLR function is defined as: \begin{equation}
     LLR_{NR} = \ln(\mathcal{L}(NR|\vec{A}(t))) - \ln(\mathcal{L}(ER|\vec{A}(t))),
\end{equation} where $\mathcal{L}$(NR$|\vec{A}$(t)) and $\mathcal{L}$(ER$|\vec{A}$(t)) are the conditional probabilities of observing nuclear and electron recoil events given a waveform $\vec{A}$(t), respectively. Values greater than zero are considered to be more likely  to be nuclear recoil events. Each of these terms use Bayes' theorem; they are defined by: 
\begin{equation}
     \ln(\mathcal{L}(NR|\vec{A}(t))) = \sum_i \ln\left(\frac{P(A(t_i)|NR(t_i)) \cdot P(NR(t_i))}{P(A(t_i)|NR(t_i)) \cdot P(NR(t_i)) + P(A(t_i)|ER(t_i)) \cdot P(ER(t_i))}\right), 
\end{equation}
and
\begin{equation}
     \ln(\mathcal{L}(ER|\vec{A}(t))) = \sum_i \ln\left(\frac{P(A(t_i)|ER(t_i)) \cdot P(ER(t_i))}{P(A(t_i)|ER(t_i)) \cdot P(ER(t_i)) + P(A(t_i)|NR(t_i)) \cdot P(NR(t_i))}\right), 
\end{equation}
where $P(NR(t_i))$ and $P(ER(t_i))$ are the prior probabilities. There is no prior information, therefore, these values are both set to 0.5 in this analysis. $P(A(t_i)|NR(t_i))$ and $P(A(t_i)|ER(t_i))$ are the conditional probability that a particular waveform amplitude is observed at time $t_i$ given that it will be classified as either a nuclear or electron recoil, respectively. To extract these probabilities, we generate a histogram of the amplitudes of the waveforms in the $i$th time bin for each tagged data set (see Figure~\ref{fig:PDF_example}). Since the shape of waveforms is energy dependent, these distributions are generated separately from data in the energy regions presented in Table \ref{table:NR_ER_stats}. Due to statistical fluctuations in these distributions, we require that each amplitude distribution is fit with either an exponentially modified Gaussian or a Voigt function, depending on the best chi-squared fit value. The area under each fit is normalised to one, to create a probability density function (PDF) from which the conditional probability can be extracted. Through this process, we generate PDFs at each time bin $t_i$ for both the nuclear and electron recoil tagged datasets. An example provided in Figure \ref{fig:PDF_example} shows the histogram of waveform amplitudes, tagged as nuclear recoils, in the $15-20$-keV energy bin for the time $t_i=$ 10~ns after pulse triggering. The fit to the amplitude distribution is also shown; in this case, an exponentially modified Gaussian was applied. From this example, the probability $P(A(t_i = 10~\text{ns})|NR(t_i = 10~\text{ns}))$ can be determined.

\begin{figure}[ht]
    \centering
    \includegraphics[scale=0.33]{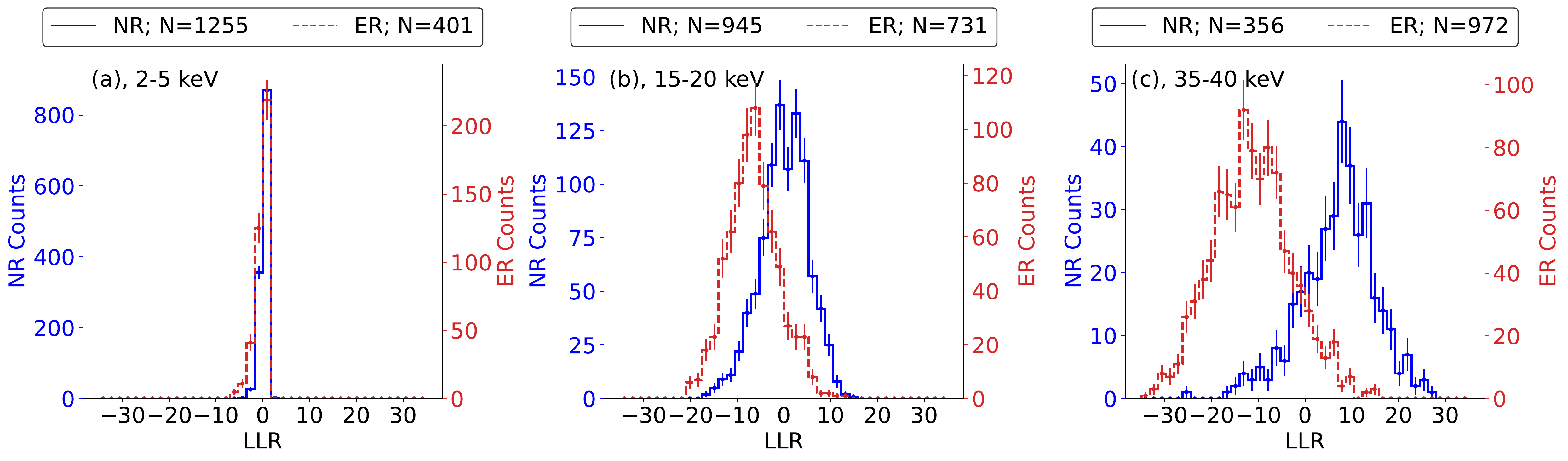}
    \caption{LLR distributions for the tagged nuclear recoil (blue) and electron recoil (red) data sets, separated into energy ranges of (a) $2-5$~keV, (b) $15-20$~keV, and (c) $35-40$~keV. Positive values denote that an event is more likely to be a nuclear recoil than an electron recoil.}
    \label{fig:LLR_dists}
\end{figure}

 Once probabilities are calculated for each sample and based on particle type, corresponding LLR values can be determined for each waveform. The LLR distributions in Figure \ref{fig:LLR_dists} show separation, similar to the lnMT approach. In the $2-5$-keV bin, both distributions are centered near zero, meaning that there is an equal likelihood that the data are from nuclear recoils or electron recoils with negative values being more electron recoil-like, and positive values being more nuclear recoil-like, under the convention we have chosen. Calculated LLR values in the higher-energy bins show the two distributions shifting further away from zero as a function of energy, which is the expected behaviour. Similar to the lnMT approach, the LLR metric has poor separation at low energies, highlighting the dependence of this method on the light-yield of the NaI:Tl crystal.

\subsection{Receiver Operator Characteristic Curves}
\begin{figure}[ht]
    \centering
    \includegraphics[width=\textwidth]{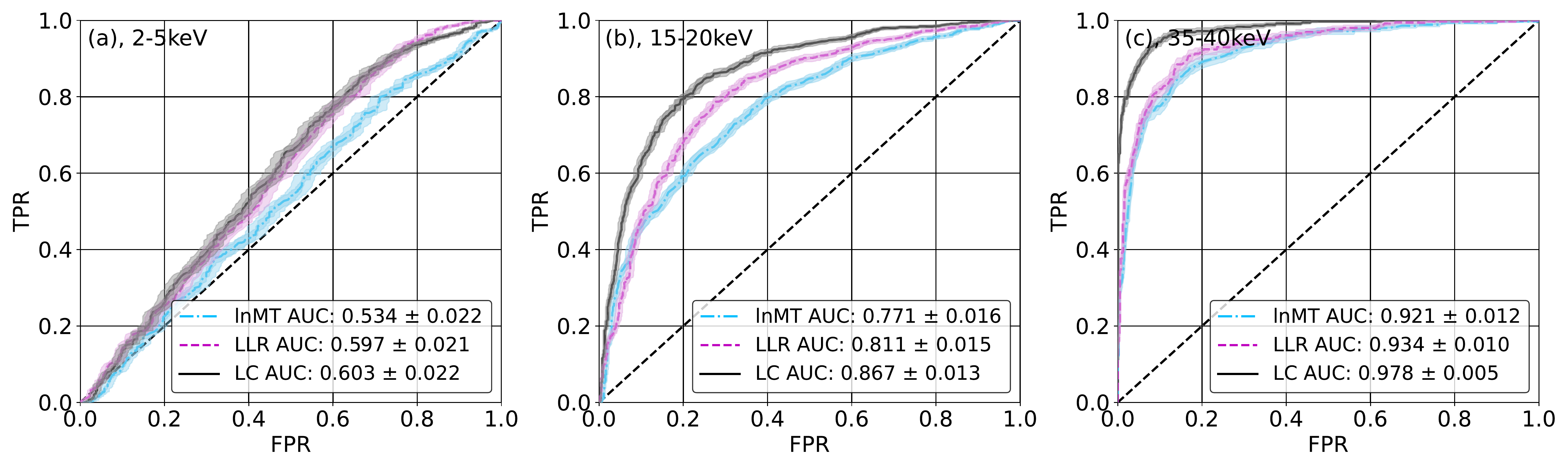}
    \caption{Receiver operator characteristic curves for all three PSD methods; lnMT (cyan), LLR (magenta) and their optimal linear combination (LC, black). Each curve is created from data binned in energy ranges of (a) $2-5$~keV, (b) $15-20$~keV, (c) $35-40$~keV with uncertainties provided to 1$\sigma$.}
    \label{fig:ROC_Curves}
\end{figure}

To gain a better understanding of the separation presented in Figures \ref{fig:lnMT_dists} and \ref{fig:LLR_dists}, it is necessary to quantify the discrimination power. This is achieved through Receiver Operator Characteristic (ROC) curves, which compare a True Positive Rate (TPR) --- a measure of correct predictions --- against a False Positive Rate (FPR) --- a measure of incorrect predictions --- to determine classifier performance~\cite{fawcett2006introduction}. We compute ROC curves for each energy bin for both approaches. These are provided in Figure \ref{fig:ROC_Curves}, along with an Area Under the Curve (AUC), which is a coarse measure to quantify the discrimination power of a given approach, with greater discrimination power corresponding to larger AUC values. The uncertainties in these ROC curves and their AUC values are determined through error propagation of the Poissonian uncertainty within the associated histograms and are provided to 1$\sigma$. We have also included the ROC curves for a linear combination of these approaches, which is discussed further in the following section. The $2-5$~keV ROC curves --- the WIMP-relevant energy range --- that are shown in Figure \ref{fig:ROC_Curves}(a) have AUC values of 0.534 $\pm$ 0.022 and 0.597 $\pm$ 0.021 for lnMT and LLR respectively. This is considered to be poor classification power, with AUC values of 0.5 --- the diagonal line --- corresponding to a random classifier. The discrimination power improves as a function of energy, with the highest energy bin, $35-40$~keV having AUC values of 0.921 $\pm$ 0.012 and 0.931 $\pm$ 0.010, for the respective techniques. Each energy bin has AUC values for the two approaches within 10\% of each other, showing that the two techniques have comparable performance, although the LLR approach has slightly better discriminatory performance in general.

\subsection{Improving discrimination with a linear combination}
\begin{figure}[ht]
    \centering
    \includegraphics[width=\textwidth]{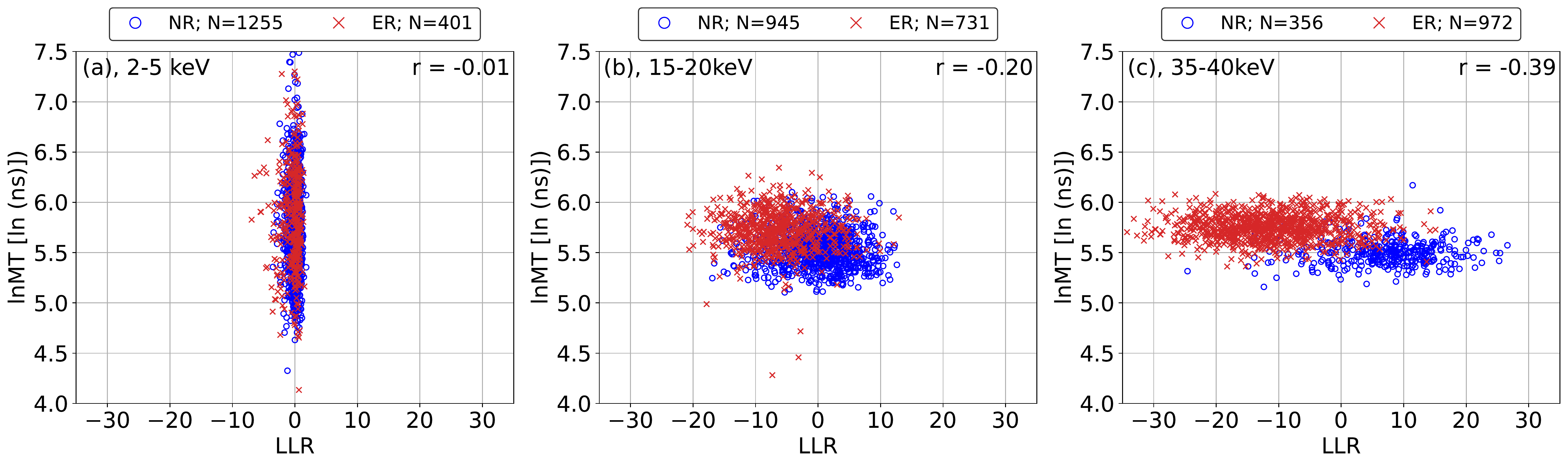}
    \caption{(a)-(c) Correlation plots: lnMT data (y-axis) versus LLR data (x-axis) for the tagged nuclear recoil (blue) and electron recoil (red) data sets, binned in energy ranges of (a) $2-5$~keV, (b) $15-20$~keV, (c) $35-40$~keV. There is a negative correlation, r, increasing in magnitude with energy, up to $r = -0.39$ for the $35-40$-keV bin.}
    \label{fig:Correlations}
\end{figure}
A key question is: can the two techniques described above be combined to further improve the discrimination power? Figure \ref{fig:Correlations} shows the correlation between the lnMT and LLR approaches. The overall data set generates a weak correlation, increasing from -0.01 in the $2-5$-keV bin up to -0.39 in the $35-40$-keV bin. This implies that the lnMT and LLR methods extract complementary classification information from the waveforms. 

\begin{figure}[ht]
\begin{subfigure}[t]{.5\textwidth}
  \centering
  \includegraphics[width=\textwidth]{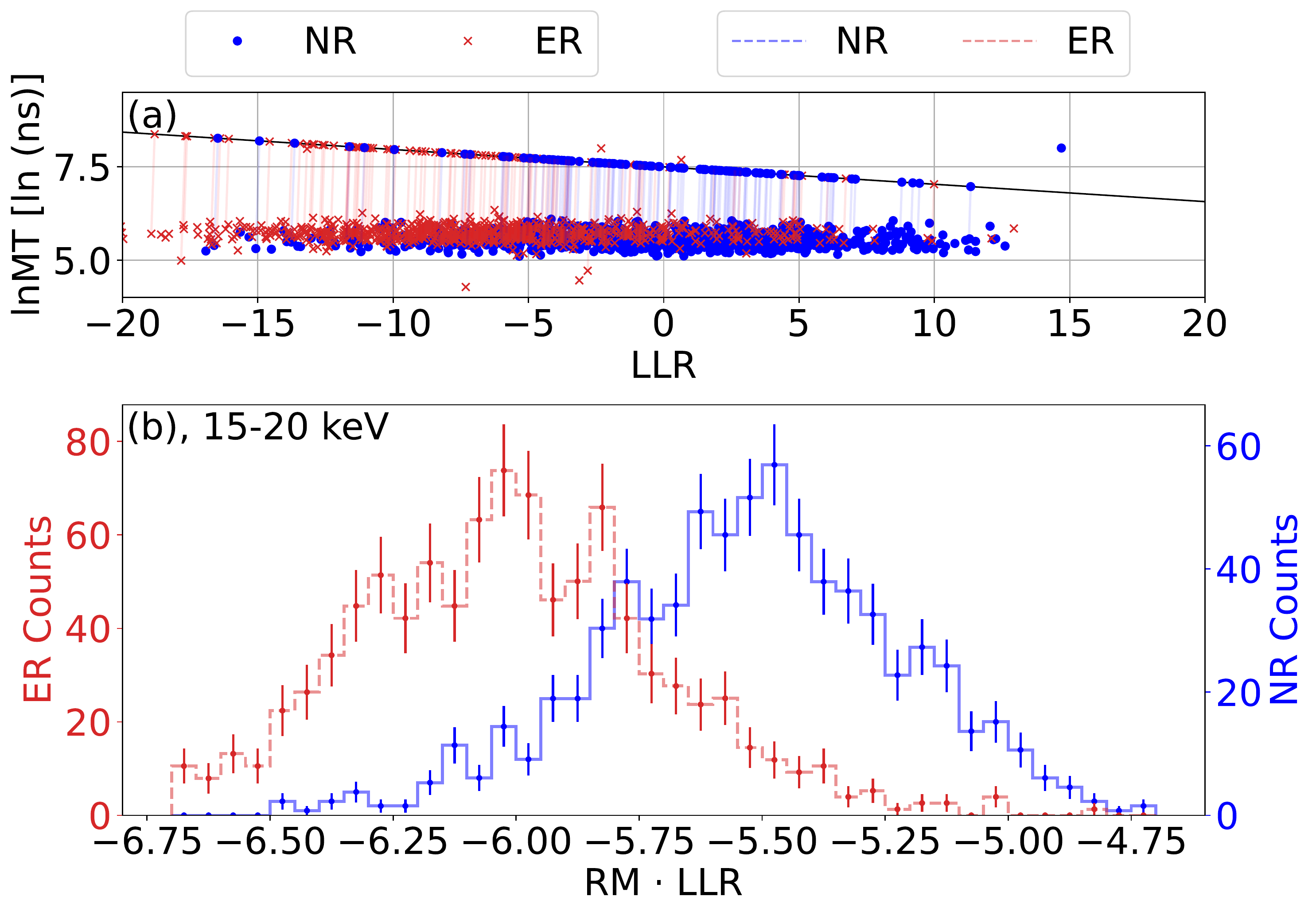}  
  \label{fig:sub-first}
\end{subfigure}
\hfill
\begin{subfigure}[t]{0.5\textwidth}
  \centering
  \includegraphics[width=1.0\linewidth]{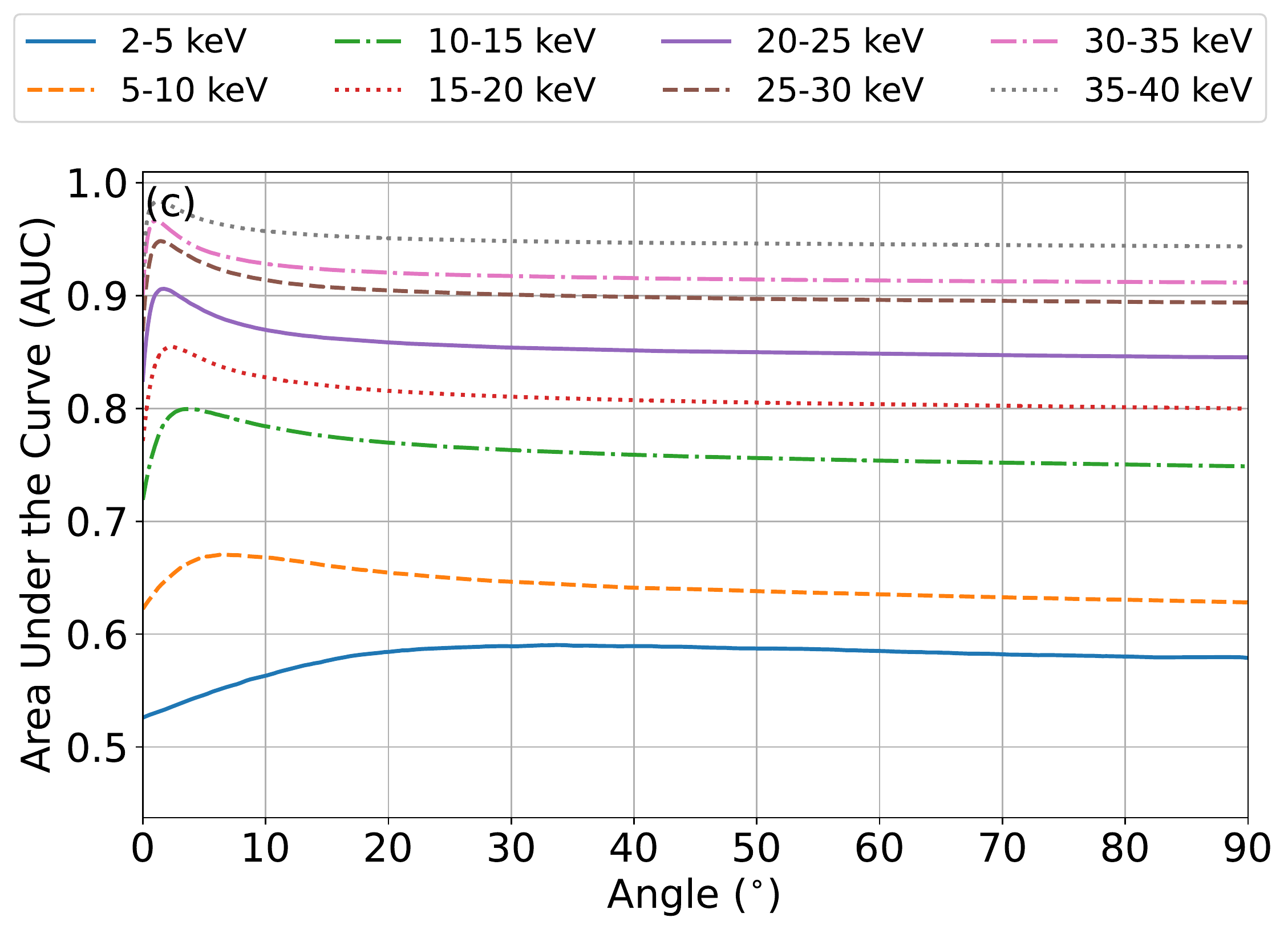}  
  \label{fig:sub-second}
\end{subfigure}
  \caption{(a) lnMT and LLR values of tagged nuclear recoil (blue) and electron recoil (red) events are projected onto a line to maximise separability. (b) The histogram associated with this projection; the projection values are determined from a dot product of LLR with a rotation matrix RM. (c) From projected histograms at various angles, ROC curves and associated AUC values are calculated. The deduced AUC values as a function of angle are shown here for each energy bin; there is a clear maximum AUC value and an associated angle.}
    \label{fig:projection}
\end{figure}
To determine a linear combination of these metrics that will give better discrimination, we project the scatter data onto a projection line (see Figures \ref{fig:projection}(a) and \ref{fig:projection}(b)). To determine the best angle of projection, the scatter data are first projected onto a line at 0$^{\circ}$ to the y-axis and a ROC curve for that projection is generated from which the AUC is deduced. The projection line is then rotated with a step size of 0.2$^{\circ}$, and ROC curves with corresponding AUC values are generated for all angles.

The AUC of each curve is then plotted as a function of the angle $\theta$, with this process repeated for each energy bin, as shown in Figure \ref{fig:projection}(c). It can be seen that almost all energy bins have distinct maximum values, corresponding to the best classification power, at small angles from 1.1$^{\circ}$ to 6.5$^{\circ}$ with decreasing energy. The $2-5$~keV bin does not produce the same shape and has a maximum AUC value at 33.5$^{\circ}$. The ROC curves with the greatest AUC values are selected as the best classification curve and are provided in Figure \ref{fig:ROC_Curves}, labelled as LC, with the lnMT and LLR curves for comparison. Each new ROC curve is determined at a different angle, because the optimal linear combination for each energy bin is different. In each case, the linear combination has a greater AUC value than either individual approach and is considered to be the best classifier of the three approaches, with an AUC value achieved in the $35-40$~keV bin of 0.969 $\pm$ 0.007.

\begin{figure}[ht]
    \centering
    \includegraphics[scale=0.4]{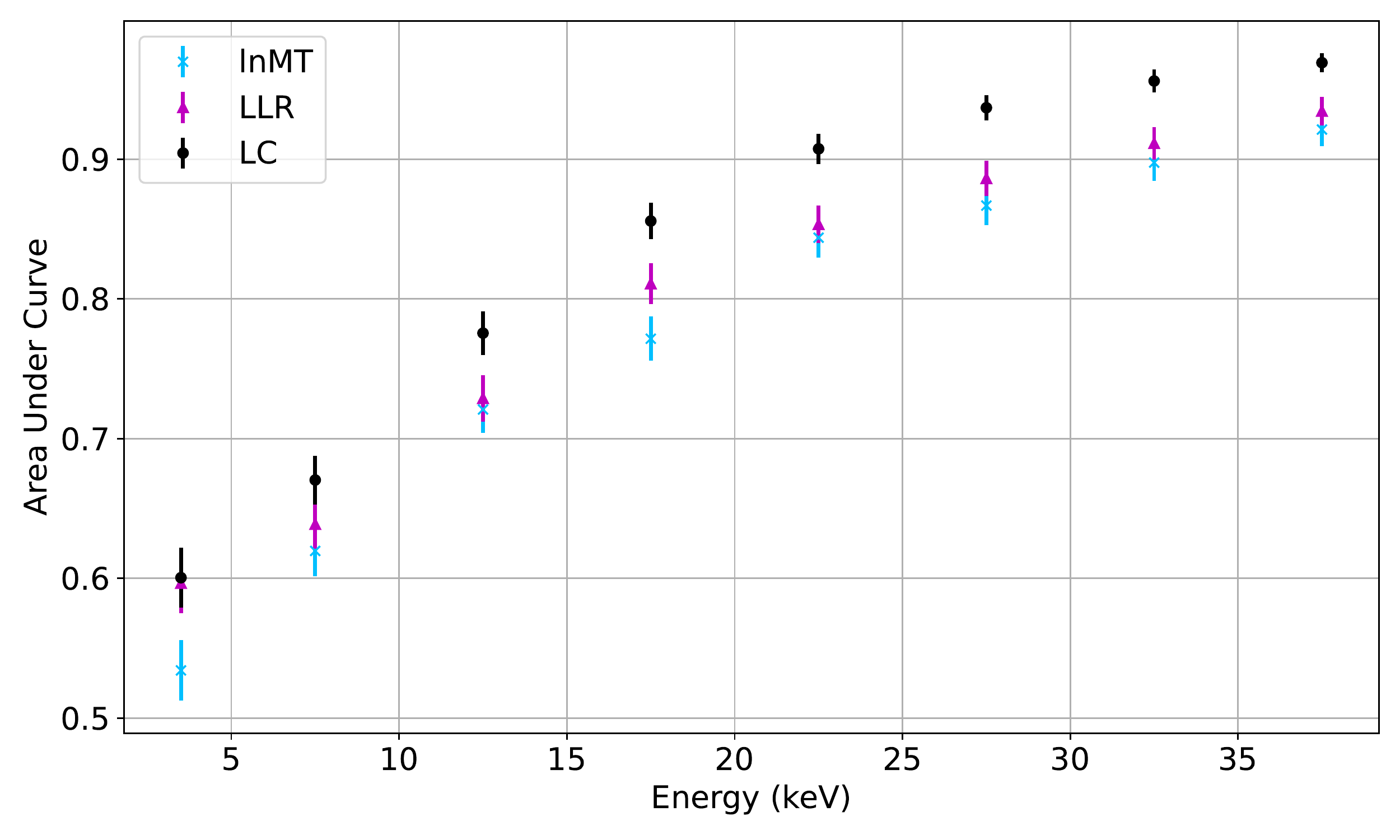}
    \caption{Areas under the curve (AUC values) for the lnMT (cyan) and LLR (magenta) approaches, along with their optimal linear combination (LC, black) are provided for all energy bins in the analysed $2-40$~keV energy range, with uncertainties provided to 1$\sigma$. The linear combination LC consistently has a greater AUC value and is considered to be the best classifier for each energy bin.}
    \label{fig:AUC_vs_energy}
\end{figure}

Figure \ref{fig:AUC_vs_energy} shows the AUC values with 1$\sigma$ uncertainty for the three approaches over all eight energy bins in the analysed $2-40$~keV energy range. In each bin, the LLR approach has better performance than the  lnMT technique (within error). This shows that, as a standalone approach the LLR is an improvement over the conventional lnMT approach. Furthermore, a linear combination of these two techniques provides improved classification power across all energies.

\section{Conclusion and Future Work}
In this work, we introduced the log-likelihood ratio (LLR) as a particle identification technique for NaI:Tl, aiming to improve on the discrimination power of the conventional charge-weighted mean-time (lnMT) approach. The LLR technique was shown to outperform the lnMT approach at all energies between 2 to 40~keV. The key result of this work is that a linear combination of the LLR and lnMT approaches has even greater performance than either individual approach. An avenue of research to be explored is classification with non-linear combinations. The data presented in Figure \ref{fig:Correlations} are not linearly separable, therefore the linear combination presented is one of many possible solutions. Support vector machines are an approach to extend this work to include non-linear combinations by constructing a linear boundary in a transformed, higher-dimensional space~\cite{yu2015neutron}. Additionally, boosted decision trees are a popular means of classification that accurately map non-linear data relationships and easily adapt to additional parameters~\cite{yang2005studies}. These approaches can then be extended to other machine learning techniques, such as Convolutional Neural Networks (CNNs), to extract the features embedded in each waveform to better discriminate between nuclear and electron recoil events. CNNs have seen use in discrimination between neutron and gamma pulse shapes previously~\cite{griffiths2020pulse}, but this approach is reliant on greater statistics for network training.

An important issue in the current analysis is the low light yield from the NaI:Tl crystal. There is an expectation that this will be improved by a factor of 5 to 10 for dark matter direct-detection experiments~\cite{adhikari2018initial}. The implication is that the separation seen here at $15-20$~keV is representative of $2-3$~keV energy events in dark matter direct-detection experiments. In future, these approaches will be applied to the SABRE experimental data, with the potential for an improvement in the SABRE sensitivity to WIMPs. With these pulse shape discrimination techniques applied, their potential impact on sensitivity to WIMPs in dark matter direct-detection experiments can be assessed.  

\section{Acknowledgements}
The authors would like to thank the technical staff at the ANU Heavy Ion Accelerator Facility for design and construction of the detector housings and beamline hardware, as well as performing detector alignment. The authors would also like to thank Alan Duffy from Swinburne University for the use of this NaI:Tl crystal. This research was supported by the Australian Research Council Centre of Excellence for Dark Matter Particle Physics (CDM, CE200100008), the Australian Research Council Discovery Program (DP170101675), and the International Technology Centre Pacific (ITC-PAC, FA520919PA138). A.A, B.J.C, J.T.H.D, M.S.M.G, T.J.G B.P.M, and M.J.Z acknowledge the support of the Australian Government Research Training Program. N.J.S, W.J.D.M, and Y.Y.Z acknowledge the support of the Australian Research Council Centre of Excellence for Dark Matter Particle Physics. National Collaborative Research Infrastructure Strategy (NCRIS) Heavy Ion Accelerators (HIA) capability for operation of the Heavy Ion Accelerator Facility is acknowledged. 
\bibliographystyle{unsrt}
\bibliography{PSD_of_LE_NR_ER_for_PID_NimA_paper.bib}
\end{document}